# HISTOGRAM EQUALIZATION OF THE IMAGE

Irem DOKEN
Department of
*Telecommunication and electronics Engineering*
Istanbul Technical University
Istanbul, Turkey
iremdoken@gmail.com

Melih GOKDEMIR
Department of
*Electrical and Electronics Engineering*
Istanbul Technical University
Istanbul, Turkey
melih.gkdmr@gmail.com

W.T AL-SHAIBANI
Department of
*Electrical and Electronics Engineering*
Istanbul Technical University
Istanbul, Turkey
tawfiqwaleed@gmail.com

Ibraheem SHAYEA
Department of
*Telecommunication and electronics Engineering*
Istanbul Technical University
Istanbul, Turkey
ibr.shayea@gmail.com

*Abstract—* **The relevance and impact of probability distributions on image processing are the subject of this study. It may be characterized as a probability distribution function of brightness for a certain area, which might be a whole picture. To generate a histogram, the probability density function of the brightness is frequently calculated by counting how many times each brightness occurs in the picture region. The brightness average is defined as the sample mean of the brightness of pixels in a certain region. The frequency is shown by the histogram. The histogram has a wide range of uses in image processing. It could, for starters, be used for picture analysis. Second, the functions of an image's brightness and contrast, as well as the final two uses of equalizing and thresholding. Normalizing a histogram is one technique to convert the intensities of discrete distributions to the probability of discrete distribution functions. The technique to equalize the histogram is to control the image's contrast by altering their intensity distribution functions. The major goal of this procedure is to give the cumulative probability function a linear trend (CDF). A method of segmentation is to divide a section of the picture into constituent areas or objects.**

## I. INTRODUCTION

The probability can be defined as the likelihood of an event occurring. The frequentist and subjectivist definitions of probability are the most prevalent. Both use a quantitative technique to calculate a probability and assign a number between 0 and 1, or any value between 0% and 100%. A number of 0 indicates that there is no chance that an event will occur. This indicates that the event will not take place. A value of one, on the other hand, indicates that the event will undoubtedly occur.

The probability is acknowledged in a frequentist perspective as the proportion of times a particular occurrence occurs in an infinite or extremely large number of ventures done in stable conditions. The ratio between the number of endeavors with a positive outcome and the total number of experiences is the relative frequency. A medical specialist, for example, may consider conducting a clinical trial to examine the entire clinical response following a certain medical therapy and determine if the outcome is favorable or negative. The ratio between the number of patients who were healed and the total number of patients who received medicine is the hypothetical frequency of response to therapy. On the other hand, according to the subjectivist viewpoint, probability is defined as a person's level of belief in the likelihood of a certain event. The Bayesian touch's distinctive concept is that all unknown quantities may be assigned a probability. To put it another way, any point of uncertainty may be expressed in probabilistic terms. The likelihood is a declaration of an assessment of the researcher's event based on the facts provided in this touch. As a result, the Bayesian system employs the Bet scheme idea to convert the belief rate into a number. Probability is defined as the price a person considers reasonable to pay in exchange for receiving a value of 1 if an event occurs or a value of 0 if the event does not occur. The level of faith a person has is determined by their ideas.

## II. METHODOLOGY

This section gives a quick overview of the relevant definitions and functions that are utilized in histogram equalization of the image. This part attempts to give the reader with a foundation of knowledge in order for them to comprehend the millstones that lie beneath the actual labor later on.

### A. Sample Space and Events

Any collection of items, known as points or elements, is referred to as a set. Space, the cosmos, or the universal set refers to the largest feasible pool of points for assessment. Theory space is referred to as the sample space in Probability. If every member of A is also an element of B, the set A is called a subset of B. If every component of A is also an element of B and there is at least one element of B that does not belong to A, A is called a proper subset of B. Equivalent sets of equal sets are two sets, A and B. The empty or null set an is defined as a set with no points. The intersection of two sets, A and B, is a set made up of their shared components. The union of two sets, A and B, produces a set that contains all points that belong to either A or B, or both. A set that comprises of all subjects is the difference between two sets, A and B

### B. Probability Space

A Probability Space is a mathematical pattern that assures that a random process or test has an official model. There are three components to a probability space. Sample space, event space, and probability function are the three. Total results are used to create sample space. The entire event space is the collection of all events. The probability function evaluates the likelihood of an occurrence. Each event possibility, according to probability theory, is a number between 0 and 1.

The conditional probability of an occurrence B is the chance that it will happen given the knowledge that an event A has already occurred. P(B|A) is the probability condition. If events A and B are independent, the conditional probability of event B given event A is simply P. (B). If events A and B are dependent, then the probability of the intersection of A and B is defined by P(A and B) = P(A)P(B|A).

$$P(A|B) = \frac{P(A,B)}{P(B)} \tag{1}$$

## C. Random Variable

A Random Variable is a variation with an unknown value or a function that assigns values to the results of each experiment. A random variable is a numerical variable whose assessed value varies from one experiment repetition to the next. Random variables are numerical assessments that have a high level of unpredictability each time they are repeated. A deterministic function called a random variable X assigns a real number to each outcome in the sample space. A continuous random variable is one with an unlimited number of potential values. It's made up of infinity of different values. A discrete random variable is one that has just one or a countably infinite number of potential values.

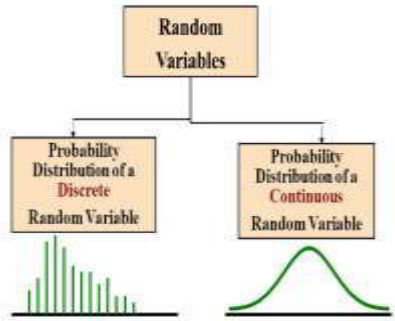

Fig. 1. Random Variables

## D. Discrete Bivariate Distribution

If X and Y are two random variables defined on the same sample space S, they can be defined in the context of the same experiment to examine how they interact. bivariate probability function could be defined as shown below.

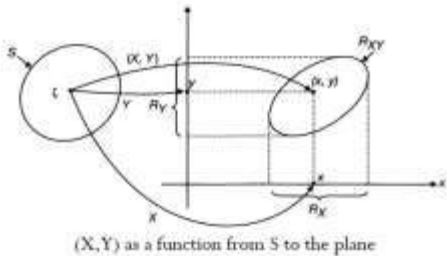

Fig. 2. Random Variable on Sample Space

$$p(x, y) = P(X = x \text{ and } Y = y) \qquad (2)$$

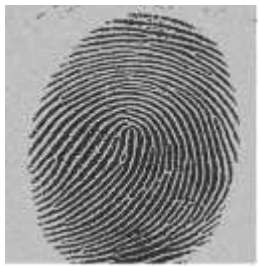

Fig. 3. Bimodal Image Histogram of fingerprint[21]

## E. Probability Mass Function

A probability mass function is a function over a discrete random variation X's sample space. It offers an absolute value to X probability. Countably nfinite is a probability mass function. Summing over a discrete probability mass function can be used to calculate probabilities. A probability mass function can be used to describe an experiment. Finite and infinite summation methods can be used to solve countably infinite issues.

## F. Probability Density Function

The probability density function is an integral function used to calculate probabilities for a continuous random variable. The probability density function graph is a curve that connects itself to the horizontal axis. The likelihood is the proportion of this region that is included between any two values. The probability density function that fits between such values describes the result of an observation. A probability density function is associated with each random variable.

## G. Probability Distribution Function

A probability distribution function is a function that describes the probability distribution. The value that the system will take on a given value or group of values is shown by a distribution function. A Probability Distribution is a collection of all potential outcomes of a random variable, together with their probability values. The Probability Distribution of a Random Variable X is a set of probabilities associated with all possible values for X. A mathematical function known as the Probability Distribution is a type of mathematical function. The probability distribution of several possible outcomes in an experiment offers the likelihood of their occurrence. The probability distribution function could be called with some different names.

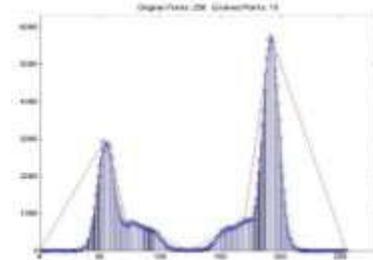

Fig. 4. Sampled Continuous Probability Distribution[21]

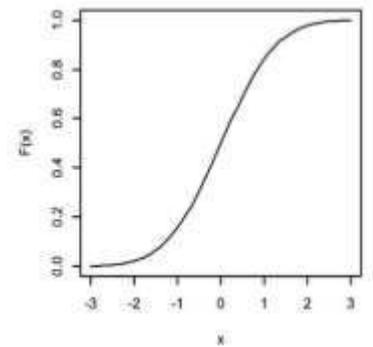

Fig. 5. Probability Distribution Function

## H. Brightness

- Probability Distribution Function of the Brightness

It may be characterized as a probability distribution function of brightness for a certain area, which might be a whole picture. P(x) is the probability distribution function, as we all know. The specified brightness value of x might be less than or equal to the brightness of the regions in this situation. While the probability of brightness has a rise from -∞ to ∞, P(x) rises from 0 to 1. P(x) specialty is monotonic and does not have a decrease, so there dP/dx≥0. Therefore;

$$p(x) \geq 0 \text{ and } \int_{-\infty}^{\infty} p(x)dx = 1 \quad (3)$$

- Probability Density Function Of The brightness

p(x) is the probability density function, while p(x) is the probability distribution function. p(x)Δx. In this area, we can calculate the likelihood of brightness changes between two points. x and x+Δx.

$$p(x)\Delta x = \left(\frac{dP(x)}{dx}\right)\Delta x \quad (4)$$

There are some functions of 2D images and their Fourier Transforms;

TABLE I. FUNCTIONS OF 2D IMAGES AND THEIR FOURIER TRANSFORMS

| Shape | Equations | |
|---|---|---|
| Rectangle | $R_{a,b}(x,y) = \frac{1}{4ab}u(a^2 - x^2)u(b^2 - y^2)$ | $\left(\frac{sin2\pi af_x}{\pi af_x}\right)\left(\frac{sin(2\pi bf_y)}{\pi bf_y}\right)$ |
| Pyramid | $R_{a,b}(x,y) \otimes R_{a,b}(x,y)$ | $\left(\frac{sin2\pi af_x}{\pi af_x}\right)\left(\frac{sin(2\pi bf_y)}{\pi bf_y}\right)$ |
| Pill Box | $P_a(r) = \frac{u(a^2-r^2)}{\pi a^2}$ | $2\left(\frac{J_1(2\pi af)}{\pi af}\right)$ |
| Cone | $P_a(r) \otimes P_a(r)$ | $\left(\frac{J_1(2\pi af)}{\pi af}\right)^2$ |
| Gaussian | $g_{2D}(r,\sigma) = \frac{1}{2\pi\sigma^2}\exp\left(-\frac{r^2}{\sigma^2}\right)$ | $G_{2D}(f,\sigma) = \exp(-2\pi^2 f^2 \sigma^2)$ |
| Peak | $\frac{1}{r}$ | $G_{2D}(f,\sigma) = \exp(-2\pi^2 f^2 \sigma^2)$ |
| Exponential Decay | $e^{-ar}$ | $\frac{2\pi a}{(w^2+a^2)^{3/2}}$ |

x is the width of the brightness gap in a picture with integer brightness amplitudes. The brightness probability density function is frequently calculated by creating a histogram, h[x], by counting how many times each brightness occurs in the picture region. The area of the underside of the histogram can be equalized by normalizing the histogram. To put it another way, if you're looking for a unique method to express

$$p[x]=\frac{1}{\Lambda}h[x] \text{ with } \Lambda = \sum_x h[x] \quad (5)$$

- Average

The sample mean of the brightnesses of the pixels in a given region is what the brightness average means. Assume that the standard is represented by $m_x$, and that the pixel area is represented by;

$$m_x=\frac{1}{\Lambda}\sum_{(m,n)\in R} x[m,n] \quad (6)$$

In addition, another formula declared as;

$$m_x=\frac{1}{\Lambda}\sum_a a \cdot h[a] \quad (7)$$

The mean of the average brightness' probability distribution function is estimated by $u_x$.

- Standard Deviation

Standard deviation is an unbiased estimation, $s_x$. R represents $s_x$ of the area of pixels is

$$s_x = \sqrt{\left(\frac{1}{\Lambda-1}\right)\sum_{(m,n)\in R}(x[m,n]-m_x)^2} \quad (8)$$

$$= \sqrt{\frac{\sum_{(m,n)\in R} x^2[m,n] - \Lambda m_x^2}{\Lambda-1}}$$

$$s_x = \sqrt{\frac{(\sum_x x^2 h[x]) - \Lambda m_x^2}{\Lambda-1}}$$

The probability distribution function of standard deviation is estimated by σ_x.

- Coefficient of Variation & Mode

Coefficient variation is dimensionless and represented by;

$$CV=\frac{s_x}{m_m}100\% \quad (9)$$

The mode is the value of the brightnesses which has the most frequent distribution. It is not confident that there will be a mode value or is unique

- Signal to Noise Ratio (SNR)

There are different definitions of signal to noise ratio, and it depends on standard deviation $s_n$. If it is assumed that signal is between $x_{min} \leq x \leq x_{max}$, so;

$$SNR=20\log_{10}\left(\frac{x_{max}-x_{min}}{s_n}\right) dB$$

Or; (10)

$$SNR=20\log_{10}\left(\frac{m_x}{s_n}\right) dB$$

$$SNR=20\log_{10}\left(\frac{s_x}{s_n}\right) dB$$

here is an example for the area specified in the image for all explanations above;

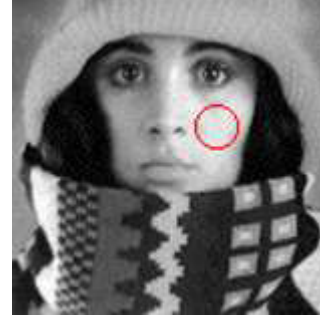

Fig. 6. Example of Values of Specifications for The Area[22]

TABLE II. EXAMPLE OF VALUES OF SPECIFICATIONS FOR THE AREA

| Statistics | Image | ROI |
|---|---|---|
| Avrage | 137.7 | 219.3 |
| Standard deivation | 49.5 | 4.0 |
| Minimum | 56 | 202 |
| Median | 141 | 220 |
| Maximam | 241 | 226 |
| Mode | 62 | 220 |
| SNR(db) | NA | 33.3 |

*I. Histograms*

The frequency is shown as a histogram. Histograms represent the intensity values of pixels in image processing. For example, the x-axis of a histogram states gray level intensities, whereas the y-axis proclaims the frequency of such intensities.

There are 256 shades of gray in a grayscale picture. As a result, the scale of our histogram graph is 0 to 255. The initial half of the first half, as seen in the graph, has a high frequency and corresponds to the darker section.

The histogram has a wide range of uses in image processing. It could, for starters, be used for picture analysis. Second, the functions of an image's brightness and contrast, as well as the final two uses of equalizing and thresholding.

1. General Histogram Specifications

Let show greyscale with r and z to be random variables with corresponding continuous probability density functions (PDF) denoted by $p_r(r)$ and $p_z(z)$. With regards to that notation, declare input and output which processed image kindly.

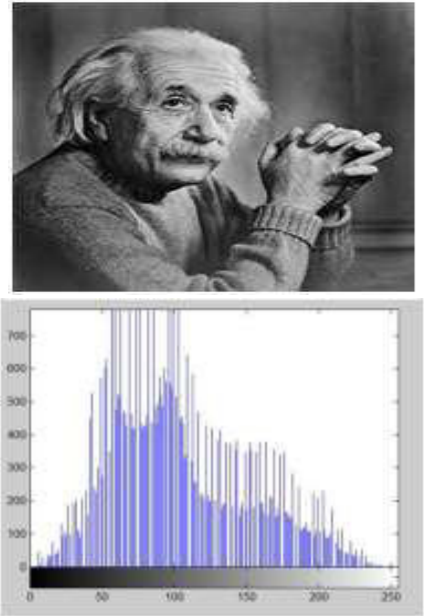

Fig. 7. Histogram Graphic of Einstein's Photograph[23]

2. Transformation of Histogram

Normalizing a histogram is one technique to convert the intensities of discrete distributions to the probability of discrete distribution functions. As a result, each histogram value is intended to be split into a pixel number. This is because a digital picture is made up of discrete values

$$n_{kn} = \frac{nk}{length \times width} = pr(rk) \tag{11}$$

3. Equalization of Histogram

To regulate contrast of the image via their intensity distribution functions' modifying is the way of equalizing the histogram. This process's main idea is to give a linear trend to the cumulative probability function (CDF).

The base of equalization of the histogram is the usage of the cumulative distribution function. The cumulative distribution function is the sum of all probabilities in the probability density function

$$cdf(x) = \sum_{k=-\infty}^{x} P(k) \tag{12}$$

A linear CDF is a uniform distribution that we want to achieve.

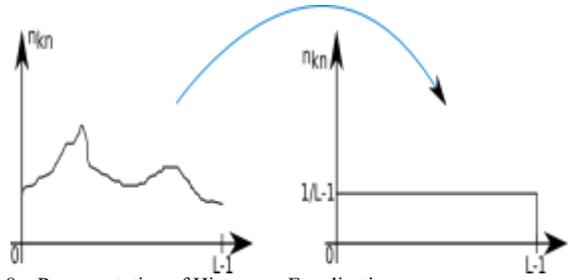

Fig. 8. Representation of Histogram Equalization

So there, the formulation is;
$$S_k = (L-1)cdf(x) \tag{13}$$

4. Brightness Preserving Bi-Histogram Equalization

$X_m$ represented to mean of image X and accepted that $X_m \in \{X_0, X_1, \ldots, X_{L-1}\}$. Therefore input image can be taken as two sub-images $X_L$ and $X_U$ as;

$$X = X_L \cup X_U$$
$$X_L = \{X(i,j) | X(i,j) \leq X_m, \forall X(i,j) \in X\} \tag{14}$$
$$X_U = \{X(i,j) | X(i,j) > X_m, \forall X(i,j) \in X\}$$

Regarding that to define their probability density functions;

$$p_L(X_k) = \frac{n_L^k}{n_L} \tag{15}$$

where k = 0, 1, …, m, and

$$p_U(X_k) = \frac{n_U^k}{n_U} \tag{16}$$

where k = m+1, m+2, …, L-1. Then so, cumulative density functions are;

$$c_L(x) = \sum_{j=0}^{k} p_L(X_j) \tag{17}$$

And

$$c_U(x) = \sum_{j=m+1}^{k} p_U(X_j) \tag{18}$$

by this way;
$$f_L(x) = X_0 + (X_m - X_0)c_L(x) \tag{19}$$

$$f_U(x) = X_{m+1} + (X_{L+1} - X_{m+1})c_U(x) \tag{20}$$

And finally, it could be expressed
$$Y = \{Y(i,j)\}$$
$$= f_L(X_L) \cup f_U(X_U) \tag{21}$$

Where

$$c_L(x) = \sum_{j=0}^{k} p_L(X_j) \tag{22}$$

$$f_L(X_L) = \{f_L(X(i,j)) | \forall X(i,j) \in X_L\} \tag{23}$$

And
$$f_U(X_U) = \{f_U(X(i,j)) | \forall X(i,j) \in X_U\} \tag{24}$$

If one note that $0 \leq c_L(x), c_U(x) \leq 1$, it is easy to see that $f_L(X_L)$ equalizes the sub-image $X_L$ over the range $(X_0, X_m)$ whereas $f_U(X_U)$ equalizes the sub-image $X_U$ over the range $(X_{m+1}, X_{L-1})$. As a consequence, the input image X is equalized over the entire dynamic range (X0, X L-1) with the constraint that the sample less than the input mean are mapped to (X0, X m), and the samples more remarkable than the mean are mapped to (X m+1, X L-1).

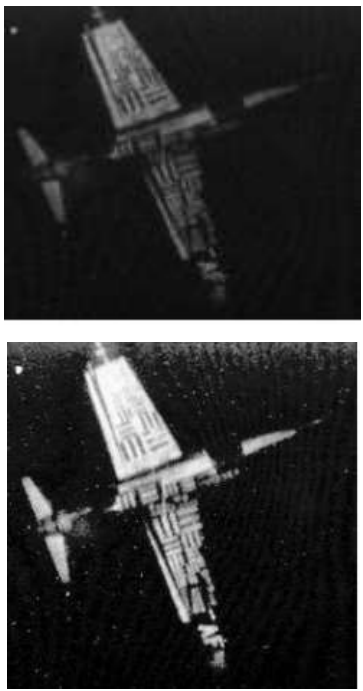

Fig. 9. Original Image and Result of BBHE[24]

Also, histogram graphs are placed below belongs to those figures;

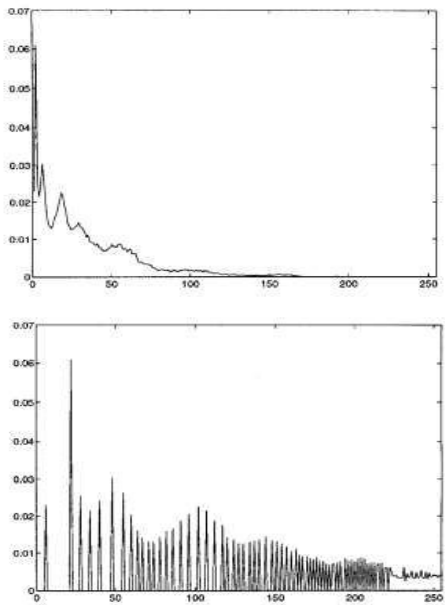

Fig. 10. Histogram Graphs of Original and BBHE

5. Segmentation with a Histogram

A method of segmentation is to divide a section of the picture into constituent areas or objects. The final success or failure of the automated analytical method is used to determine the accusation of segmentation.

The methods of segmentation require two essential properties: discontinuity and similarity of intensity levels. Portioning an image based on sudden variations in intensity is one of the types (e.g., edges of the image).

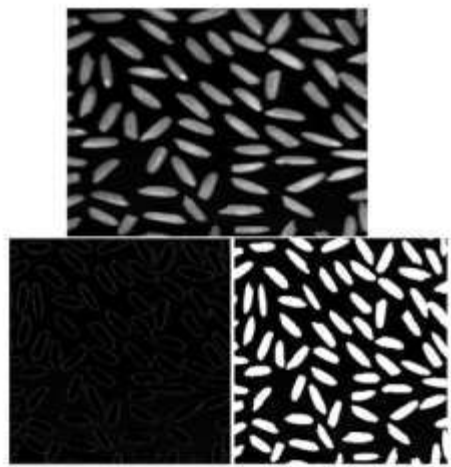

Fig. 11. . Bimodal Gray Scale Image and After Image Segmentation[21]

III. MATERIAL & METHOD

A. Matlab

Matlab is a programming environment designed for engineers and scientists. Matlab's language is built on the matrix, which allows for natural mathematical statements.

Matlab analyzes data, develops an algorithm, and simulates the probability grayscale histogram in image processing in this study.

Pictures are generally stored in 8 bits of data for accuracy. As a result, integer values ranging from 0 to 255 must be assigned. The color of each red, green, and blue (RGB) pixel is indicated by a value between 0 and 255. The benefits include a universal language that may be used to create a variety of problem-solving solutions. It uses algorithms that are both resilient and sophisticated. It's also simple to use probability distributions to analyze the signal. Matlab's database and computer vision techniques are expanding. It's simple and quick to use. It already has an image processing toolkit installed, as well as the ability to integrate other libraries like OpenCV.

A license, on the other hand, is required to use Matlab. Most institutions, on the other hand, give students with a Matlab license for academic purposes. The processing in this study is done using a CT scan.

IV. CONCULUSIONS

In this paper, an algorithm is used to estimate the parameters of some probability density function using a medical image. Here both the program and image structure has been considered. The program shows the probability function using the image, for basically, code was designed to probability function properties and image process structures. Firstly, the algorithm changes of image structure and validate for the availability of image processing. Second, the algorithm creates the matrix system that is important because the Matlab program works based on matrix structures.Thirdly the algorithm applies image process properties and probability function features.

The simulation results show that the algorithm closely characterized the multi-modal distributed data of probability function. So it can play a role in medical monitoring and prediction. This paper focused on the probability distribution of the medical image process. Results show that the available structure of the system can improve the medical image system by applying the probability distribution function. Further works could provide accurate high-level results.

ACKNOWLEDGMENT

The authors appreciate the reviewers' valuable time to review this work.